\documentclass[twocolumn,showpacs,preprintnumbers]{revtex4}

\usepackage{graphicx}
\usepackage{dcolumn}
\usepackage{bm}
                                                                                
\begin{document}
\draft

\title{
Robustness of d-Density Wave Order to Nonmagnetic Impurities
}
\author{
Amit Ghosal$^{1}$ and Hae-Young Kee$^{2}$
}
\affiliation{$^{1}$Department of Physics and Astronomy, McMaster University,
Hamilton, Ontario Canada L8S 4M1 \\
$^{2}$ Department of Physics, University of Toronto,
Toronto, Ontario Canada M5S 1A7
}
\date{\today}

\begin{abstract}
Effect of finite density of nonmagnetic impurities on a coexisting phase of
d-density wave (DDW) order and d-wave superconducting (DSC) order is studied
using Bogoliubov-de Gennes (BdG) method. The spatial variation of the
inhomogeneous DDW order due to impurities has a strong correlation with that
of density, which is very different from that of DSC order. The length scale
associated with DDW is found to be of the order of a lattice spacing. The
nontrivial inhomogeneities are shown to make DDW order much more robust to
the impurities, while DSC order becomes very sensitive to them. The effect of
disorder on the density of states is also discussed.
\end{abstract}
\maketitle


\paragraph{Introduction}
One of the recent proposals 
in the context of high temperature cuprates is that a true broken
symmetry state dubbed as d-density wave state (DDW) is 
responsible for the pseudogap phenomena.\cite{sudip}
This phase
was first suggested in relation to the excitonic insulators\cite{rice},
and it was found as one of the ground states of
the t-J type  model.\cite{brad}
The DDW is a particle hole condensate with angular momentum 2. 
The ordered state can be characterized by the circulating current
arranged in an alternating pattern on a square lattice,
which can be detected as a Bragg scattering signal 
in neutron scattering measurements.\cite{brad2,hykee}
But the neutron scattering experiments\cite{mook,keimer,buyers}
in cuprates remain controversial.
Thus, the definite conclusion on the relevance of the DDW order
to the cuprates
requires more precise
experiments on various doping concentration of cuprates,
and further theoretical studies on the properties of this new order.
Especially, the effect of the nonmagnetic
impurities on DDW order is an important subject to investigate, 
since any well-prepared cuprate sample contain an intrinsic
disorder, minimally from non-stoichiometry.

The simplest possible description of the impurity effect
is the self-consistent T-matrix approximation (SCTMA)
\cite{maki2}.
This mean field picture excludes 
not only the freedom of the ordered patterns, but also
the interference of the impurities.
Within this approximation, the thermodynamics were found
to be identical to those of a d-wave BCS superconductor (DSC) in
the unitary limit.\cite{maki}  From the density of states, one can see
that electrons are localized close to the Fermi energy, and the change in
the transition temperature is given by the Abrikosov-Gorkov
formula known in BCS superconductors.\cite{maki}
Within the standard non-crossing approximation,  
the similarity between the DDW and DSC is based on
the d-wave symmetry of the gap.

In this paper, we study
the effect of impurities on DDW order and for the case where DDW coexists
with DSC using Bogoliubov-de Gennes (BdG) technique. 
This method is the mean field approximation,  
but it allows spatial inhomogeneity in order parameter.
In the case of  the disordered DSC with a short  coherence length,
it was shown that the superfluid stiffness
is significantly larger than that predicted by the SCTMA, due
to the nontrivial spatial structures of the order parameter\cite{ghosald}.

We found that the DDW order is more robust than
the DSC order to the impurities, which cannot be understood within
the conventional T-matrix approach.
The physical ground for our findings will be discussed  
later.


\paragraph{Model}
We model two dimensional disordered DSC and DDW order by the 
following Hamiltonian.
\begin{eqnarray}
{\cal H} &=& 
 -t\sum_{<ij>,\alpha} (c_{i\alpha}^{\dag} c_{j\alpha} + h.c.)
+  \sum_i \left(V(i)-\mu \right) n_i
\nonumber\\
 &  & \hspace{-0.5cm}
 + J\sum_{<ij>}\left({\bf S}_i \cdot {\bf S}_j
- n_i n_j /4 \right) + W\sum_{<ij>,\alpha,\beta} n_{i \alpha}
n_{j \beta}.
\end{eqnarray}
The first term is the kinetic energy
which describes electrons, with spin $\alpha$ at site $i$ created by
$c_{i\alpha}^{\dag}$, hopping between nearest-neighbors $<ij>$ on a
square lattice.  
The disorder potential $V(i)$ in the second term is an independent random
variable at each site which is either $+ V_0$, with a
probability $n_{\rm imp}$ (impurity concentration), or zero,
and  $\mu$ is the chemical potential.
The last, interaction term \cite{footnote1}
is chosen to lead to a coexisting  DSC and DDW order
ground state in the disorder-free system, where
${\bf S}_i $ and $n_i$ are the spin and density operators, respectively.

The mean field decomposition of the above Hamiltonian leads to
following BdG equations \cite{pgdg,ghosals}.
\begin{equation}
\left(\matrix{\hat\xi & \hat\Delta \cr \hat\Delta^{*} & -\hat\xi^{*}} \right)
\left(\matrix{u_{n} \cr v_{n}} \right) = E_{n}
\left(\matrix{u_{n} \cr v_{n}} \right),
\label {eq:bdg}
\end{equation}
where
$\hat\xi u_{n}(j) = -\sum_{\delta} \{t + \Psi(j;\delta) e^{-i {\bf Q.r_j}} \}
u_{n}(j+\delta) + (V(j)-\tilde{\mu}_j)u_{n}(j)$ and $\hat\Delta u_{n}(j)
= \sum_{\delta}\Delta(j+\delta;\delta) u_{n}(j+\delta)$,
and similarly for $v_{n}(j)$.
The local DSC pairing ($\Delta$) and DDW ($\chi = {\rm Im} \Psi$)
 amplitudes on a bond $(j;\delta)$ 
are defined by
\begin{eqnarray}
\Delta(j;\delta) &=& -\frac{J+W}{4}\langle c_{j+\delta \downarrow}
c_{j \uparrow} + c_{j \downarrow}c_{j+\delta \uparrow}\rangle,
\nonumber\\
\Psi(j; \delta) & =&
\frac{ J + 2W}{4}\langle c^{\dag}_{j+\delta \alpha}c_{j \alpha}
- c_{j \alpha} c^{\dag}_{j+\delta \alpha}\rangle e^{-i {\bf Q.r_j}},
\end{eqnarray}
where
$\delta = \pm{\hat{\bf x}}, \pm{\hat{\bf y}}$. 
The inhomogeneous Hartee and Fock shifts are given by $\tilde{\mu}_j =
\mu + (\frac{J}{4}+W) \sum_{\delta}\langle n_{j+\delta} \rangle$
and $Re[\Psi(j;\delta)]$ respectively.

We numerically solve for the BdG eigenvalues $E_n \ge 0$ and eigenvectors
$\left(u_{n},v_{n}\right)$ on a lattice of $N$ sites with periodic boundary
conditions. We then calculate the d-wave pairing amplitude
$\Delta(j;\delta) = (J+W)\sum_n\left[u_n(j+\delta)v^*_n(j) 
+ u_n(j)v_n^*(j+\delta)\right]/4$ 
and the DDW order and Fock shift as
the imaginary and real parts of $\Psi(j;\delta)
 = (J+2W)
\sum_n\left[v^*_n(j)v_n(j+\delta)-u_n(j)u^*_n(j+\delta)\right]/4$ 
at $T=0$,
and the density $\langle n_j \rangle = 2\sum_n |v_n(j)|^2$.
These are fed back into the BdG equation, and the process iterated
until self consistency \cite{footnote3} is achieved for {\it each}
of the (local) variables defined on the sites and bonds of the
lattice. The chemical potential $\mu$ is chosen to obtain a given
average density $\langle n \rangle = \sum_i \langle n_i \rangle/N$.
We define the site dependent order parameters in terms of the bond
variables as, $\Delta(j) = \left[\Delta(j;+\hat{x})-\Delta(j;+\hat{y})
+ \Delta(j;-\hat{x}) - \Delta(j;-\hat{y}) \right]/4$ and similarly for
$\chi(j)$.

We have studied the model at $T=0$ for a range of parameters and lattice
sizes up to $40\times 40$. Here we focus on $J = 1.16$, and $W = 0.6$, in
units of $t = 1$, with $\langle n \rangle=0.95$ on systems of typical size
$30\times 30$.
For these parameters, and $n_{\rm imp} = 0$, the maximum DSC gap is
$\Delta_{{\rm max}}=0.16$ and the maximum DDW gap is $\chi_{{\rm max}}=0.31$.
In the pure system our calculations reproduce a phase diagram of
$\Delta_{{\rm max}}$ and $\chi_{{\rm max}}$ as functions of filling
similar to Ref. \cite{zhu}.
For the impurity potential we choose $V_0 = 100$, close to the unitary limit.
The results are averaged over 10 different realizations of the random
potential.
\begin{figure}
\scalebox{.38}{\includegraphics{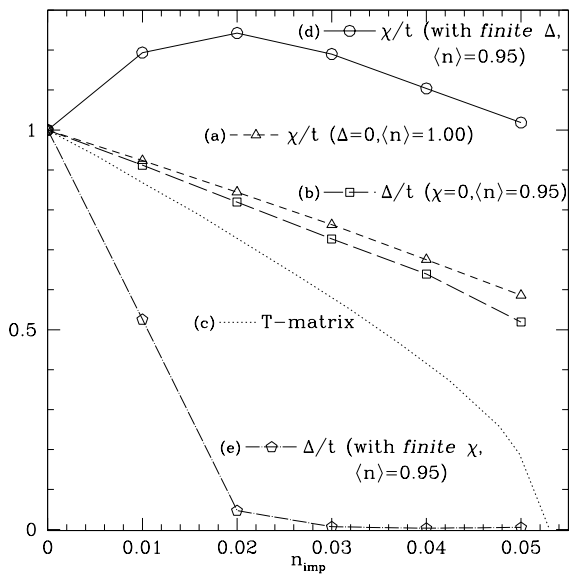} }
\caption{
The evolution of $\Delta$ and $\chi$ (normalized by their pure values)
with $n_{\rm imp}$ is plotted under different conditions (see text).
}
\label{fig:order}
\end{figure}


\paragraph{Effect of Impurity on DDW and DSC Orders}
We summarize our main results in Fig. (1), where we plot the disorder
dependence of different orders (normalized to $n_{\rm imp}=0$ values).
Let us first look at the line (a) that represents the behavior of $\chi$
as a function of $n_{\rm imp}$ at half filling ($\langle n \rangle =1$).
At this filling DDW is the stable order and DSC order in fact vanishes for
the pure system. Comparing the behavior of $\chi$ with the
results from
SCTMA calculations (represented by (c) curve) we see that the DDW order
is more robust to impurities than predicted by SCTMA. 

On the other hand,
away from half filling when we force $\chi=0$ in BdG equations, DSC becomes
the surviving order and the $n_{\rm imp}$ dependence of $\Delta$ is given by
the (b) line, which in fact is very similar to $\chi$ in the (a) curve. Such
robustness of the DSC order to impurity had been studied before \cite{ghosald},
and it is attributed primarily to the fact that -- each impurity affects
superconductivity rather inhomogeneously by destroying SC order within a
small region (of size determined by coherence length $\xi$) around it. 
Hence the long range order is not globally affected. We find from our
current numerical results that similar picture holds for DDW order as well,
and $\chi$ is also affected locally by impurity, keeping long range DDW
order robust. 

However, similar study for the coexisting phase of DSC + DDW order at
$\langle n \rangle=0.95$ reveals surprisingly that, 
superconducting order  is severely affected by disorder
(curve (e)) in the coexisting phase, much more so than in the absence of
$\chi$. 
On the contrary, the DDW order (curve (d))
coexisting with DSC order 
becomes even more robust to impurities.
In fact for low $n_{\rm imp}$, $\chi$ even increases with impurity.
The rest of the paper is organized towards the detailed understanding
of these unexpected results.

From Fig. (1d) we saw that the DDW order increases for small $n_{\rm imp}$.
To get a further insight, we study
the spatial structures of the order parameter 
(particularly at large $n_{\rm imp}$)
on the lattice for each impurity configuration. In Fig. (2a) we present a
Grey-scale plot of $\chi$ on a
typical $30\times 30$ lattice at $n_{\rm imp}=0.06$ for a given realization
of scatterers. The dark (light) regions represent larger (smaller) values of
$\chi$. Comparing this structure with Fig. (2b), that gives the spatial
distribution of $|\langle n \rangle -1 |$ for the same $n_{\rm imp}$,
we see that $\chi$ is large in
space where {\rm local} density is close to 1 (half filling). 
The strong spatial correlation between these two panels is striking,
although it is not exact;
the scale of modulation of $\chi$ is somewhat larger than that of density.
However, the strong tie of local $\langle n \rangle$ and $\chi$ suggest that
the length scale of fluctuation of $\chi$ would be governed by that of
$\langle n \rangle$, which is rather small (of the order of $k_F^{-1}$).
This can be understood as follows.

The length scale associated with the DDW order,  
$\xi_{\rm DDW} \sim 1/\chi$. 
When impurity is introduced, the bond current attached to the impurity
site is forced to be zero. So does the density.
However, the bond current 
"near" the impurity site is re-constructed
to satisfy the current conservation, and one should note that
the healing length is of order of a lattice spacing.
How the magnitude of the re-constructed bond-current
is determined?
This magnitude is strongly related to the local
density.
The electron density depletes close to
impurities and increases at locations far from it,
to keep the average at the desired value. 
Since at low disorder, a large number of sites
attain $\langle n_i \rangle \sim 1$,  $\chi$  increases at those sites;
the DDW order is most stable near half filling, where perfect
nesting occurs for our model. 
As a result average $\chi$ increases. At very large $n_{\rm imp}$, local
density would be either much larger or smaller than 1, and $\chi$
would decrease everywhere. This argument can be substantiated
by looking into our results for each configuration of impurities.

For $\langle n \rangle = 1$, introduction of impurity
makes local density only to deviate from half filling. As a result $\chi$
decreases monotonically as found in Fig. (1a). The above argument for the
behavior of DDW order with impurity is independent of the coexisting DSC
order and we also found similar trend in $\chi$ as in Fig. (1d) for
$\langle n \rangle < 1$ even in the absence of DSC order, which is consistent
with our picture. This shows that DDW order responds to the density
fluctuations due to impurities. 
On the contrary, the DSC order in the presence of impurities is {\em not}
related to the local density fluctuations as DDW is, even though
the length scale, $\xi_{\rm DSC} \sim 1/\Delta$, which is of the order
of a few lattice spacing for high temperature superconductors under
considerations. The behavior of $\Delta$ in the presence of impurities
is shown to be related to the electron-hole mixing in the real space
\cite{ghosals}; $\Delta$ is large when the density is close to the
chemical potential. 
\begin{figure}
\scalebox{.36}{\includegraphics{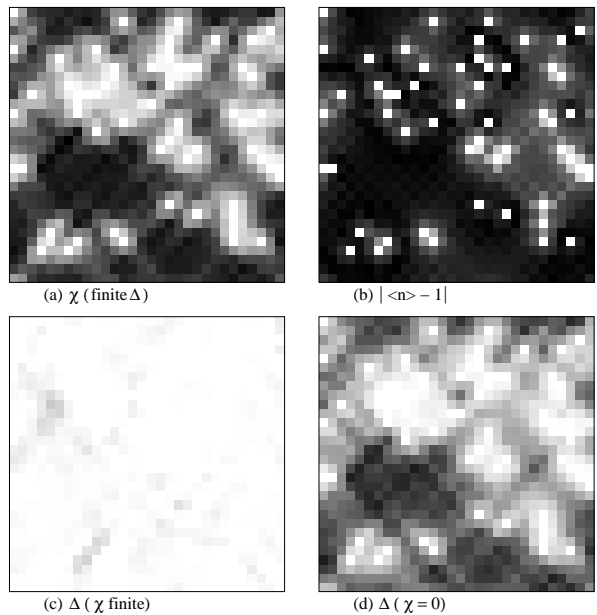} }
\caption{
Grey scale plot of DDW and DSC order on lattice for $n_{\rm imp}=0.06$ and
for a particular configuration of unitary impurities. The dark (light) region
indicates large (small) values of the variables on given locations. Panel
(a) and (b) are the $\chi$ and $|\langle n \rangle -1 |$ respectively for
a system with coexisting DDW + DSC order at $\langle n \rangle = 0.95$. 
 $\Delta$ is shown in (c) with a  finite $\chi$,
and  (d) with  $\chi$ forced to zero everywhere.
}
\label{fig:order}
\end{figure}

Fig. (2c) and (2d) presents the spatial structure of $\Delta$ on lattice
with the same $n_{imp}$ configuration in the presence and absence of
DDW order. We clearly see that the DSC is 
strongly suppressed by the impurities when coexisting with DDW
order (as also observed in Fig. (1b) and (1e)). 
The existence of DDW strongly affect the strength of the DSC,
because away from the impurities there are regions where the density
is near half-filling, hence the DDW becomes strong. Strong DDW allows
significant weight of ($\pi$,$\pi$) scattering that mixes the $+$ and $-$
lobes of the DSC order and thereby DSC becomes weak. The regions of
small density does not contribute to DSC order as well, due to the absence
of enough electrons for pairing! Thus in the inhomogeneous coexistence phase
DSC order is suppressed everywhere.
 

\paragraph{Averaged Density of States}
Let us now study the (impurity) averaged density of states (DOS)
$N(\omega) = {1 \over N}\sum_{n,i} \left[ |u_n(i)|^2\delta(\omega - E_n)
+ |v_n(i)|^2\delta(\omega + E_n) \right]$ (where we broaden the delta 
functions with a width comparable to average level spacing).
To obtain a better statistics for $N(\omega)$, we used ``Repeated Zone Scheme",
\cite{ghosalv} that describes a large effective system made out of
$10\times 10$ unit cells, each of which is of dimension $30\times 30$.
\begin{figure}
\scalebox{.38}{\includegraphics{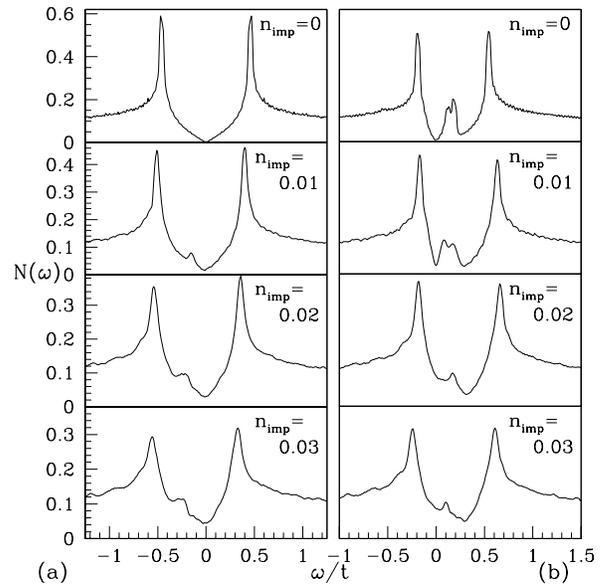} }
\caption{
Panel (a): Disorder averaged density of states $N(\omega)$ on a system
with {\em only} DDW order (at $\langle n \rangle=1$) on $10\times10$ unit
cells each of which are of size $N=30\times30$. 
It shows the robustness of DDW order (see text).
Panel (b): Similar to Panel (a), but with {\em coexisting} DDW + DSC phases
(at $\langle n \rangle=0.95$). With increasing $n_{\rm imp}$ DSC gap is
washed out, however, strong DDW order persists.
}
\label{fig:dos}
\end{figure}
In Fig. (3) we plot $N(\omega)$ as a function of $\omega$ for different
$n_{\rm imp}$ for the case of only DDW order (Panel a) and coexisting
DDW + DSC order (Panel b). For the pure system with only DDW order,
$N(\omega)$ is the standard d-wave DOS. With increasing $n_{\rm imp}$
we see that the gap-edge singularities get rounded off and a small
accumulation of states is produced at the particle side of spectrum
close to $\omega=0$. 
The accumulation of
electrons around a single impurity  effectively provide impurity screening,
which  will produce enhanced states at the
particle side of the spectrum.\cite{morr}
Such resonances from each impurity contribute to the average $N(\omega)$
and produce a broad band which is reflected as a bump in Fig. (3a). 
However, the strength of the DDW order is not
affected much (given by the relative location of the two coherence
peaks).
At this point, we should emphasize that the DOS structure for impure DSC state is
very different, where coherence peaks get strongly suppressed and a thin
gap persists at $\omega=0$ \cite{ghosald,peter}, so that $N(0)=0$ for all
$n_{\rm imp}$. From our results with DDW order, we find that
$N(0) \propto n_{\rm imp}$, which is in disagreement with the prediction of
T-Matrix result ($N(0) \propto \sqrt{n_{\rm imp}}$)\cite{maki2,maki}.

In Fig. (3b), for coexisting DSC + DDW, a double-gap DOS is expected at
$n_{\rm imp}=0$ \cite{wonkee}; superconducting gap at
$\omega=0$ and d-density wave gap at $\omega=\tilde{\mu}$. With increasing
$n_{\rm imp}$, DSC gap gets washed out and by $n_{\rm imp}=0.03$, $N(\omega)$
looks very similar for Fig. (3a) and (3b) (The overall shift for the later
case is due to the particle-hole asymmetry). This demonstrates in a different
way our main result, that, the DSC order is very sensitive to impurity
whereas DDW order is robust in the coexisting phase.


\paragraph{Summary and Discussion}
We studied the effect of  nonmagnetic impurity on DDW ordered state
using BdG technique.
While the standard SCTMA indicates that the  effect of  impurity on DDW
is similar to that on DSC, 
we found that the  spatial variation of the DDW order
has a strong correlation with that of density [Fig. (2a) and (2b)], 
and it  is 
very different from that of DSC order [Fig. (2a) and (2c)].
We discussed that this occurs because
the length scale associated with the DDW order
is of order of a lattice spacing ($\sim 1/k_F$),
which suggests that the spatial variation of DDW order is related to
the density fluctuation, while
 the DSC order is related to particle-hole mixing. 
Therefore, the effect of impurity on  the DDW order 
 is very different from that of DSC order,
which can not be obtained from the standard SCTMA method.

When  DSC and DDW coexist,
it turns out that DDW order do not care about the existence of
DSC and it still follows
the density profile in the presence of impurity.
However, DSC order would vanish almost everywhere [See Fig. (2c)].
This is because
in the region of larger density
it is killed by DDW, and in the region
of smaller density it is destroyed by disorder. 
Thus in the inhomogeneous media both DDW and impurity are acting 
to suppress the DSC order.

Our current picture brings out the unexpected results and
their understanding at the mean field level; if the DDW phase exists
in cuprates, the Bragg signal would be detected
in  neutron scattering measurements
even in the presence of strong nonmagnetic impurity, while
the  width of the Bragg peaks depends on strength of impurity.
However, the definite answer for its relevance to the cuprates
requires the understanding of the role of strong correlation,
 and interplay between different competing orders, 
which warrants further studies.


{\bf Acknowledgments}: We would like to thank Y. B. Kim and
A. Vishwanath for illuminating discussions. 
We acknowledge SHARCNet
computational facilities at McMaster University where most of the
calculations were carried out.
This work is supported by SHARCNet fellowship(AG),
NSERC of Canada(HYK), Canada Research Chair(HYK),
and Canadian Institute for Advanced Research(HYK).



\begin{references}
\bibitem{footnote1}
Note that either of the ``spin exchange" term or the ``extended Hubbard"
term of ${\cal H}_{\rm int}$ could produce both the DSC or the DDW orders.
However we keep both terms for stabilizing DDW phase (See Ref. 
 \cite{zhu}) and also to be able to tune both the orders independently.
\end{references}
\end{document}